\documentclass[preprintnumbers]{revtex4}

\usepackage[dvipdfmx]{graphicx,color}
\usepackage{epsf}
\usepackage{amsmath,amssymb}
\usepackage{bm}
\usepackage{color}
\usepackage{textcomp}
\newcommand{\bvec}{\boldsymbol}
\begin{document}
\title{Analysis of delocalization of clusters in linear-chain $\alpha$-cluster states with entanglement entropy}
\author{Yoshiko Kanada-En'yo}
\affiliation{Department of Physics, Kyoto University, Kyoto 606-8502, Japan}

\begin{abstract}
I investigate entanglement entropy of one dimension (1D) cluster states to discuss the delocalization of 
clusters in linear-chain $3\alpha$- and $4\alpha$-cluster states. 
In analysis of entanglement entropy of 1D Tohsaki-Horiuchi-Schuck-R\"opke (THSR)
and Brink-Bloch cluster wave functions, I show clear differences in the entanglement entropy 
between localized cluster wave functions and delocalized cluster wave functions.
In order to clarify spatial regions where the entanglement entropy is generated by the delocalization of clusters,    
I  analyze the spatial distribution of entanglement entropy.
In the linear-chain $3\alpha$ cluster state, the delocalization occurs dominantly in a low-density tail region while
it is relatively suppressed in an inner region because of Pauli blocking effect between clusters. In the linear-chain 4$\alpha$ state
having a larger system size than the linear-chain $3\alpha$ state, 
the delocalization occurs in the whole system. 
The entanglement entropy is found to be a measure of the delocalization of clusters in the 1D cluster systems.
 \end{abstract}
\maketitle

\section{Introduction}
A variety of cluster states have been known in light nuclei, such as
$2\alpha$ states in $^8$Be, $^{16}{\rm O}+\alpha$ states 
in $^{20}$Ne, and $3\alpha$ in 
$^{12}$C (for example, see Ref.~\cite{Fujiwara-supp} and references therein). 
In this decade, a new concept of cluster states has been proposed
to understand cluster motion in these states \cite{Tohsaki:2001an,Funaki:2002fn,Funaki:2003af,Funaki:2009zz,Zhou:2012zz,Zhou:2013ala}. That is a dilute cluster gas state where clusters are not spatially 
localized in certain positions but they are rather freely 
moving like a gas. To describe the non-localized (delocalized) cluster states, 
a new type of cluster wave function, 
so-called ``Tohsaki-Horiuchi-Schuck-R\"opke'' (THSR) wave function, has been introduced. 
The THSR wave function is essentially based on $\alpha$ clusters in a common Gaussian orbit 
having a range of the system size. It has been shown that the $2\alpha$ state for $^8$Be($0^+_1$) and 
the $3\alpha$ state for $^{12}$C($0^+_2$) can be well described by 
the THSR wave functions with a large Gaussian range 
compared with the cluster size, and therefore, these states are interpreted
as gas-like cluster states of $2\alpha$ and $3\alpha$. 
Recently, Zhou {\it et al.} have extended the THSR wave function to apply it to $^{20}$Ne
system, and shown that  the THSR wave function
can also describe the $^{16}$O+$\alpha$ states in $^{20}$Ne \cite{Zhou:2012zz,Zhou:2013ala}.

More recently, Suhara {\it et al.} have proposed that 
this concept of the $\alpha$-cluster gas is applicable also to one dimension (1D) cluster motion
in  linear-chain $n\alpha$ structures \cite{Suhara:2013csa}.
Existence of the linear-chain $n\alpha$ states has been a long standing problem. 
In the early stage, Morinaga proposed a linear-chain $3\alpha$ structure in 
$^{12}$C($0^+_2$) \cite{Morinaga:1956zza}, but this assignment was excluded from the experimental data of 
its $\alpha$ decay width \cite{suzuki72}. The possibility of the linear-chain $3\alpha$ structure
in a higher $0^+$ state is now on discussion. Negative results for the linear-chain state were obtained
by $3\alpha$-cluster models \cite{Fujiwara-supp,uegaki1}, whereas appearance of a chain-like $3\alpha$ state 
with an open triangle configuration was predicted by microscopic approaches
with no cluster assumption and by an approach with the cluster breaking effect \cite{KanadaEn'yo:1998rf,KanadaEn'yo:2006ze,Chernykh:2007zz,Suhara:2014wua} and also by an {\it ab initio} calculation
\cite{Epelbaum:2012qn}.
 Also linear-chain $4\alpha$ structures in $^{16}$O have been attracting a 
great interest, and searching for linear-chain states have been performed 
in experimental and theoretical works in this line \cite{Chevallier:1967zz,aberg82,bauhoff84,itagaki95,Freer:2004bi,Ichikawa:2011iz}. 

The conventional picture for the linear-chain $n\alpha$ structures is 
spatially localized $\alpha$ clusters arranged in 1D with certain 
intervals. Such a localized $n\alpha$ state is described by a single 
Brink-Bloch (BB) wave function \cite{brink66}. The essential result shown by Suhara {\it et al.} is that 
the wave functions of the linear-chain $3\alpha$ and $4\alpha$ states 
are described by superposing a large
number of BB wave functions, and surprisingly, they have large overlaps with 
1D-THSR wave functions \cite{Suhara:2013csa}.
This result indicates that the delocalization occurs in 1D $\alpha$-cluster motion and 
the linear-chain states can be regarded as 
1D $\alpha$ cluster gases of delocalized clusters rather than localized cluster states. 
However, as shown in Ref.~\cite{Suhara:2013csa}, the density distributions of the linear-chain 
states of $3\alpha$ and $4\alpha$
have three- and four-peak structures, respectively, which indicates partial localization of  $\alpha$ clusters
because of the Pauli blocking (repulsion) between $\alpha$ clusters. 
Even though the model in Ref.~\cite{Suhara:2013csa} was restricted in 1D configurations
and the stability of the linear-chain states against bending motion and 
$\alpha$ decays have yet to be investigated for a conclusive answer to existence of the
linear-chain states in realistic nuclear systems, their work provides the new picture 
of 1D cluster states, which is important to understand cluster phenomena 
in nuclear many-body systems, and also academically interesting.

In light nuclei, clusters are often formed by spatial correlations of constituent nucleons. 
Even in a mean-field picture, spatially correlated nucleons in a cluster 
can be described by a product of localized single-particle wave functions. 
However, such a localized cluster usually has much kinetic energy for the localization of 
center of mass motion (c.m.m.) of the cluster. If there is no potential nor Pauli blocking effect
between clusters, it is naively expected that delocalization of the c.m.m. 
of clusters occurs to release the kinetic energy. 
An ideal case of the delocalization limit is a zero-momentum cluster gas state.
The delocalization of clusters in realistic cluster states 
may depend on competition 
between kinetic energy gain and energy loss by other effects
such as potential and Pauli blocking between clusters.

To distinguish between localized and delocalized cluster structures in microscopic wave functions,
one should carefully consider the antisymmetrization of nucleons between clusters, which strongly affects inter-cluster motion at a small distance. 
The antisymmetrization effect suppresses an amplitude of the inter-cluster wave function at a small inter-cluster distance
and works as the Pauli repulsion (blocking) between clusters, 
whereas the effect vanishes at a large distance. 
If the system size is comparable to the cluster size, clusters can not move freely 
because of the Pauli blocking effect.
Therefore, the delocalization of clusters likely occurs 
not in high-density regions but in low-density regions. 
Indeed, for two-body cluster states of $2\alpha$ and $^{16}{\rm O}+\alpha$ in
$^8$Be and $^{20}$Ne, the author has shown that the delocalization of $\alpha$ clusters 
occurs at a long tail part of the inter-cluster motion \cite{Kanada-En'yo:2014vva}. 
Also in the result of the linear-chain  $3\alpha$ and $4\alpha$  states shown by Suhara {\it et al.}, 
the peak structure of density distributions shows 
the partial localization of $\alpha$ clusters in an inner region as mentioned previously. 
It may suggest the possibility that the delocalization does not occur in the whole region of the system.
I should stress here that the THSR model wave functions, which can successfully describe cluster gas states, have characters of localization and/or delocalization of clusters depending on the system size relative
to the cluster size.
In the case that the system size is large enough compared with the cluster size, 
the THSR wave function actually describe a dilute cluster gas. However, 
if the system size is as small as the cluster size,
the THSR wave function become equivalent to a localized cluster wave function
because of the antisymmetrization. 
A general question is how one can understand the delocalization of clusters in an intermediate case 
between both limits of localization and delocalization. 
To clarify the region where the delocalization occurs, one may encounter a difficulty from the antisymmetrization, i.e., Pauli blocking effect.  
The Pauli blocking effect often makes it difficult to tell the difference 
between localization and delocalization of clusters in a relatively high-density region, where 
the delocalization is usually suppressed.
Note that one can not obtain a definite answer from 
a wave function without the antisymmetrization, which 
may contain unphysical forbidden states. 

My aim is to analyze the delocalization of clusters in microscopic wave functions
with an approach free from the antisymmetrization effect.
For this aim I propose a method of analysis using entanglement entropy defined by the one-body 
density matrix. The entanglement entropy has been introduced 
by Bennett {\it et al.} in 1996 \cite{bennett96}, and widely
used in various fields such as condensed matter and 
quantum field theory (see for example 
Refs.~\cite{Calabrese:2004eu,amico08,Nishioka:2009un} 
and references therein). 
The entanglement entropy indicates how particles are entangled 
with other particles, and can be an indicator to measure 
many-body correlations as applied in 
in various fields such as condensed matter physics and quantum physics. 
For a system of independent Fermions, the wave function is given by a Slater 
determinant of single-particle wave functions, and the entanglement entropy completely vanishes. The entanglement entropy is generated by 
many-body correlations beyond a Slater determinant. In the case of cluster states, the entanglement entropy is zero in a single BB wave function for the localized cluster limit, 
and it is generated by the delocalization of clusters. 

In this paper, I analyze the entanglement entropy in 1D cluster states for $n\alpha$ and $^{16}{\rm O}+\alpha$ systems. 
Based on the analysis of the entanglement entropy, I investigate the delocalization of clusters in the
linear-chain $3\alpha$ and $4\alpha$ states predicted by Suhara {\it et al.} as well as
the $2\alpha$ state in $^8$Be($0^+_1$) and the $^{16}{\rm O}+\alpha$ states in $^{20}$Ne($0^+_1$) and  $^{20}$Ne($1^-_1$).

This paper is organized as follows. 
I describe the formulation 
in Section \ref{sec:formulation}.
Section \ref{sec:results} discusses the entanglement entropy of 1D cluster states.
The paper concludes with a summary in section \ref{sec:summary}.

\section{Formulation}\label{sec:formulation}
\subsection{Wave functions for linear-chain $n\alpha$-cluster structure}
I consider the linear-chain  $n\alpha$-cluster structures
aligned to the $z$ axis ($n$ is the number of $\alpha$ clusters). 
For simplicity, the angular momentum projection is 
not taken into account, and only 1D configurations of $n$ $\alpha$ clusters in 
intrinsic wave functions are considered in the present analysis.
It means that the (de)localization of $\alpha$ clusters 
are defined for $\alpha$-cluster motion along the $z$ axis. 
In this section, I first describe a general form of $\alpha$-cluster 
wave functions for the linear-chain structures. More details of practical model wave functions 
used in the present paper are described in the latter part of this section. 

\subsubsection{BB wave function}
I use the BB wave function \cite{brink66} for a localized $n\alpha$-cluster 
wave function,
\begin{eqnarray}\label{eq:BB}
\Phi^{n\alpha}_{\rm BB}(\bvec{R}_1,\ldots,\bvec{R}_{n})&=&\frac{1}{\sqrt{A!}}
{\cal A}\left[ \psi^\alpha_{\bvec{R}_1}\cdots  \psi^\alpha_{\bvec{R}_{n}}\right],\\
\psi^\alpha_{\bvec{R}_i}&=&\phi^{0s}_{\bvec{R}_i}\chi_{p\uparrow}
\phi^{0s}_{\bvec{R}_i}\chi_{p\downarrow}\phi^{0s}_{\bvec{R}_i}\chi_{n\uparrow}
\phi^{0s}_{\bvec{R}_i}\chi_{n\downarrow},\\
\phi^{0s}&=&(\pi b^2)^{-3/4}\exp\left[ -\frac{1}{2b^2}(\bvec{r}-\bvec{R}_i)^2\right].
\end{eqnarray}
$\psi^\alpha_{\bvec{R}_i}$ is the four-nucleon wave function of the 
$i$th $\alpha$ cluster expressed by the $(0s)^4$ harmonic oscillator (ho) shell-model configuration 
localized around the spatial position $\bvec{R}_i$. Note that a single 
BB wave function for an $n\alpha$ system is written by a Slater determinant of single-particle wave functions.
For a linear-chain structure aligned to the $z$ axis, the position parameter $\bvec{R}_i$
is set to be $\bvec{R}_i=(0,0,R_i)$, and the 1D BB wave function is expressed as
$\Phi^{n\alpha}_{\rm BB}(R_1,\ldots,R_{n})$.
 The parameter $b$ for the $\alpha$-cluster size
is chosen to be $b=1.376$ fm same as in Ref.~\cite{Suhara:2013csa}. 
General wave functions for 1D $n\alpha$ systems 
can be written by linear combination of  BB wave functions
$\Phi^{n\alpha}_{\rm BB}(R_1,\ldots,R_{n})$.

\subsubsection{1D-THSR wave function}
As shown in Ref.~\cite{Suhara:2013csa}, the linear-chain $3\alpha$- and $4\alpha$-cluster states in $^{12}$C and $^{16}$O systems 
are well described 
by the 1D-THSR wave functions proposed by Suhara {\it et al.}. The 1D-THSR wave function is 
given by linear combination of  BB wave functions with a Gaussian weight and it is in principle written as
\begin{eqnarray}\label{eq:1D-THSR}
\Phi^{n\alpha}_\textrm{1D-THSR}(\beta)=\int dR_1 \cdots dR_{n}\exp\left\{ 
-\sum^{n}_{i=1} \frac{R^2_i}{\beta^2}
\right\}\Phi^{n\alpha}_{\rm BB}(R_1,\ldots,R_{n}),\nonumber\\
\propto {\cal A} \left[
\prod_{i=1}^{n}\exp\left\{-\frac{2X^2_{ix}}{b^2}
-\frac{2X^2_{iy}}{b^2}-\frac{X^2_{iz}}{\beta^2+b^2/2}
\right\}\phi(\alpha_i)
\right],
\end{eqnarray}
where $\bvec{X}_{i}$ is the center of mass (c.m.) coordinate of the $i$th $\alpha$ cluster
and $\phi(\alpha_i)$ is the intrinsic wave function of the $\alpha$ cluster.
If the antisymmetrization is ignored, the $\Phi^{n\alpha}_\textrm{1D-THSR}(\beta)$ expresses the 
$n\alpha$ state where all $\alpha$ clusters are confined in the 
$x$ and $y$ directions in the size $b/\sqrt{2}$ while they move in the $z$ direction in the 
Gaussian orbit specified by the size parameter $\beta$, which corresponds to the system size of the linear-chain state.
When $\beta$ is large enough compared with the $\alpha$-cluster size, 
the 1D-THSR wave function describes a dilute  linear-chain gas where $n$ $\alpha$ 
clusters move almost freely like a gas in the $z$ direction.
In the present calculation, the $R_i$ integration is approximated by 
summation on mesh points in a finite box. The details are described later.

The BB and 1D-THSR wave functions 
contain the total c.m.m.
In the present paper, the c.m.m. is not removed exactly
because separation of the c.m. coordinate and intrinsic coordinates 
is technically difficult in calculation of the density matrix.
In analysis of entropy in a system of $1\alpha$ (one $\alpha$) and that of an $\alpha$ cluster with a core,  
I keep the total c.m.m. as defined in the original form.
For $2\alpha$, 3$\alpha$, and 4$\alpha$ systems, I make a correction of 
the c.m.m. to eliminate a possible artifact from $\beta$ dependence 
in the c.m.m. as explained later. 

\subsection{Entanglement entropy}
I briefly describe entanglement entropy defined by 
the one-body density matrix. More detailed explanations are given in appendixes.

\subsubsection{Density matrix}
For a wave function $|\Psi^{(A)}\rangle$ for an $A$-nucleon state, 
the one-body density matrix $\rho^{(1)}$ is defined 
in the coordinate space as
\begin{equation}
\rho^{(1)}(\bvec{r}\sigma;\bvec{r}'\sigma')=\langle\Psi^{(A)}|a^\dagger(\bvec{r}'\sigma')
a(\bvec{r}\sigma) |\Psi^{(A)} \rangle, 
\end{equation} 
where $a^\dagger(\bvec{r}\sigma)$ and $a(\bvec{r}\sigma)$ are 
creation and annihilation operators of a nucleon at the position $\bvec{r}$ 
with spin-isospin $\sigma=p\uparrow,p\downarrow,n\uparrow,n\downarrow$.
The one-body density matrix is regarded as the matrix element of the density 
operator $\hat \rho^{(1)}_{\Psi}$ for the wave function $\Psi^{(A)}$, 
\begin{equation}
\rho^{(1)}(\bvec{r}\sigma;\bvec{r}'\sigma')=\langle \bvec{r}\sigma|\hat \rho^{(1)}_{\Psi}|
\bvec{r}'\sigma'\rangle.
\end{equation} 
The diagonal element of the density matrix $\rho^{(1)}(\bvec{r}\sigma)=
\rho^{(1)}(\bvec{r}\sigma;\bvec{r}\sigma)$ indicates the one-body density of 
$\sigma$ nucleons at $\bvec{r}$.
Using the orthonormal single-particle bases $\{|l\rangle \}$ that diagonalize   
the density matrix,  the density operator and density matrix are written as  
\begin{eqnarray}
\hat \rho^{(1)}_{\Psi}&=&\sum_{l}|l\rangle \rho^{(1)}_l \langle l|, \\
\rho^{(1)}(\bvec{r}\sigma;\bvec{r}'\sigma')
&=&\sum_{l} \langle\bvec{r}\sigma|l\rangle
\rho^{(1)}_{l}\langle l| \bvec{r}'\sigma' \rangle, \nonumber\\
&=&\sum_{l} \phi_l(\bvec{r}\sigma)
\rho^{(1)}_l\phi^*_l(\bvec{r}'\sigma'),
\end{eqnarray} 
where $\phi_l(\bvec{r}\sigma)=\langle \bvec{r}\sigma|l\rangle$ is the wave function 
for the single-particle state $|l\rangle$, and
\begin{eqnarray}
\rho^{(1)}_l&=&\langle\Psi^{(A)}|a^\dagger_l a_l  |\Psi^{(A)}\rangle, \\
0&\leq& \rho^{(1)}_l \leq 1, 
\end{eqnarray}
is the eigen value of the density matrix and indicates the occupation probability of the single-particle state $|l\rangle$.
The trace of the density matrix $\rho^{(1)}$ equals to the particle number as
\begin{eqnarray}\label{eq:particle-number}
A&=&{\rm Tr} \rho^{(1)}=\sum_l\rho^{(1)}_l=\sum_\sigma \int\rho^{(1)}(\bvec{r}\sigma) d \bvec{r}\\
\rho^{(1)}(\bvec{r}\sigma)&=&\sum_l \rho^{(1)}_l\phi^*_l(\bvec{r}\sigma)\phi_l(\bvec{r}\sigma), 
\end{eqnarray}

\subsubsection{Entanglement entropy}
The entanglement entropy 
is defined by the one-body density matrix as,
\begin{equation}
S^{(1)}=-{\rm Tr}\rho^{(1)}\log \rho^{(1)}=- \sum_l \rho^{(1)}_l \log \rho^{(1)}_l.
\end{equation}
The entanglement entropy is zero if 
a wave function $|\Psi^{(A)}\rangle$ is a Slater determinant, because 
$\rho^{(1)}_l=1$ for occupied states and $\rho^{(1)}_l=0$ for unoccupied states.
Thai is, the density operator $\hat \rho^{(1)}_{\Psi}$
satisfies $\{\hat \rho^{(1)}_{\Psi}\}^2=\hat \rho^{(1)}_{\Psi}$
in the single-particle Hilbert space for a Slater determinant wave function \cite{ring-schuck}.

In analogy to the expression 
\eqref{eq:particle-number} for the particle number by the $\sigma$ sum and 
$\bvec{r}$ integral of the ``local'' density $\rho^{(1)}(\bvec{r}\sigma)$, 
I define ``local" entanglement entropy as follows, 
\begin{eqnarray}
S^{(1)}&=&  \sum_\sigma \int   s^{(1)}(\bvec{r}\sigma) d\bvec{r},\\
s^{(1)}(\bvec{r}\sigma)&=&\sum_l  \left[-\rho^{(1)}_l\log\rho^{(1)}_l\right]
 \phi^*_l(\bvec{r}\sigma)\phi_l(\bvec{r}\sigma).
\end{eqnarray}
Here the factor $[-\rho^{(1)}_l\log\rho^{(1)}_l]$ is contribution 
of the single-particle state $|l\rangle$ in  $S^{(1)}$, and $\phi^*_l(\bvec{r}\sigma)\phi_l(\bvec{r}\sigma)$ means the 
density distribution in $|l\rangle$ and it is normalized as $\sum_\sigma\int d\bvec{r}\phi^*_l(\bvec{r}\sigma)\phi_l(\bvec{r}\sigma)=1$.
Therefore, the local entanglement entropy $s^{(1)}(\bvec{r}\sigma)$ reflects 
spatial distributions of the important states $|l\rangle$ that contribute to 
the total entanglement entropy, whereas it is hardly affected  by almost occupied states 
having $\rho_l^{(1)}\approx 1$. 

\subsubsection{Entanglement entropy for linear-chain $\alpha$-cluster states}

For a system of independent Fermions, the wave function is given by a Slater 
determinant and it has zero entanglement entropy, $S^{(1)}=0$. 
In general, the entanglement entropy indicates how particles are entangled with other particles, 
and can be an indicator for many-body correlations.

In a $n\alpha$-cluster state, 
the total wave function $\Psi$ is spin and isospin symmetric and the one-body density 
is block diagonal with respect to $\sigma$;
$\rho^{(1)}(\bvec{r}\sigma;\bvec{r}'\sigma')) =\rho^{(1)}(\bvec{r};\bvec{r}')\delta_{\sigma\sigma'}$, where the reduced matrix $\rho^{(1)}(\bvec{r};\bvec{r}')$ is independent to $\sigma$.
Therefore, I can discuss the density matrix and the entanglement entropy
with the reduced matrix in the subspace, 
\begin{eqnarray}
A&=&4n,\\
n&=&\int\rho(\bvec{r}) d \bvec{r},\\
\rho(\bvec{r}) &=&\rho^{(1)}(\bvec{r}\sigma,\bvec{r}\sigma),\\
S^{(1)}&=&4S,\\
S&=&\int  s(\bvec{r})  d\bvec{r},\\
s(\bvec{r})&=&  s^{(1)}(\bvec{r}\sigma).
\end{eqnarray}
In the present paper, $\rho(\bvec{r})$, $S$, and $s(\bvec{r})$ 
indicate the density, the total entanglement entropy, and the local entanglement entropy, respectively,  
defined by the reduced density matrix.

In the present 1D cluster wave functions,
single-particle wave functions for $x$ and $y$ coordinates 
are common for all nucleons and give no contribution to the entanglement entropy.
Therefore, I discuss only $z$ dependence of the local density and the local entanglement 
by integrating $x$ and $y$ coordinates as
\begin{eqnarray}
\rho(z)&=&\int  \rho(\bvec{r}) dx dy,\\
s(z)&=&\int  s(\bvec{r}) dx dy.
\end{eqnarray}

\subsection{Model wave functions for $n\alpha$, $^8$Be, and  $^{20}$Ne systems} 
\subsubsection{Wave functions for linear-chain 2$\alpha$, 3$\alpha$, and 4$\alpha$ states}
For 1D-THSR wave functions of linear-chain 2$\alpha$, 3$\alpha$, and 4$\alpha$ states, 
the $R_i$ integration is approximated by 
summation on mesh points at 1 fm intervals. 
To eliminate the $\beta$ dependence of 
the c.m.m., I correct the c.m.m. by shifting the 
cluster position, $R_i\rightarrow R'_i=R_i-R_G$ of basis BB wave functions 
($\Phi^{n\alpha}_{\rm BB}$) in $\Phi^{n\alpha}_\textrm{1D-THSR}(\beta)$ with $R_G \equiv (R_1+\cdots+R_{n})/n$.
Consequently, the 1D-THSR wave function is approximated as  
\begin{eqnarray}
\Phi^{n\alpha}_\textrm{1D-THSR}(\beta)\rightarrow\sum_{R_1=0,\pm 1, \ldots} \cdots\sum_{R_{n}=0,\pm 1, \ldots}
\exp\left\{ 
-\sum^{n}_{i=1} \frac{R^2_i}{\beta^2}
\right\}\Phi^{n\alpha}_{\rm BB}(R'_1,\ldots,R'_{n}),
\end{eqnarray}
where mesh points of the summation are truncated in a finite box
$|R'_i|\le 12$ fm. The correction of c.m.m. is equivalent to replacing
the $\beta$ dependent c.m.m. $\Phi_G(\beta)$ in the original 1D-THSR 
with $\Phi_G(0)$ localized at the origin to eliminate the $\beta$ dependence in the c.m.m.,
\begin{eqnarray}
\Phi^{n\alpha}_\textrm{1D-THSR}&=&\Phi^{z}_G(\beta)\Phi_{\rm int}(\beta)\rightarrow 
\Phi^{z}_G(0)\Phi_{\rm int}(\beta),\\
\Phi^{z}_G(\beta)&\propto&\exp\left\{-\frac{A}{2b^2}X_{Gx}^2 -\frac{A}{2b^2}X_{Gy}^2-\frac{A}{4\beta^2+2b^2}X_{Gz}^2\right\}
\end{eqnarray}
where $\bvec{X}_G$ is the total c.m. coordinate.

\subsubsection{Intrinsic wave function for $^{8}$Be($0^+$)}
For $^8$Be($0^+$), I use $2\alpha$-cluster wave functions.
Funaki {\it et al.} have shown that $^8$Be$(0^+_1)$ is well described by the 
three dimension  (3D) THSR wave function of 2$\alpha$ \cite{Funaki:2002fn,Funaki:2009zz}.
In the present paper, I consider the spherical 3D-THSR wave function, which can have 98\% overlap with the 
exact solution of the 2$\alpha$ state for $^8$Be$(0^+_1)$ \cite{Kanada-En'yo:2014vva}.
The definition of the 3D-THSR wave functions is explained in Appendix \ref{app:3D-THSR}.

I use the $R^2$-weighted 1D-THSR wave function of $2\alpha$, 
\begin{equation}\label{eq:int-8Be}
\Phi^{\rm int}_{^8{\rm Be}}(R^2;\beta)=\int dR R^2
\exp\left\{ 
-\frac{R^2}{2\beta^2}
\right\}\Phi^{2\alpha}_{\rm BB}(R_1=+R/2,R_2=-R/2),
\end{equation}
which is regarded as the intrinsic state of the spherical 
3D-THSR wave function before the angular momentum projection as
\begin{eqnarray}
P^{J=0}\Phi^{\rm int}_{^8{\rm Be}}(R^2;\beta)\approx
\int d\bvec{R} 
\exp\left\{ 
-\frac{\bvec{R}^2}{2\beta^2}
\right\}\Phi^{2\alpha}_{\rm BB}(\bvec{R}_1=+\frac{\bvec{R}}{2},\bvec{R}_2=-\frac{\bvec{R}}{2})
=\frac{\Phi_G(0)}{\Phi_G(\beta)}\Phi^{2\alpha}_\textrm{3D-THSR}(\beta) ,
\end{eqnarray}
where the factor ${\Phi_G(0)}/{\Phi_G(\beta)}$ is the correction of the $\beta$ dependent 
c.m.m., and $\Phi_{\textrm{3D-THSR}}(\beta)$ is the 
spherical 3D-THSR wave function, which are described in Appendix \ref{app:3D-THSR}.
In the practical calculation, the $R$ integration in Eq.~\eqref{eq:int-8Be} is approximated by 
summation of mesh points at , ${R=0,1,\ldots,24 \ {\rm fm}}$. 

\subsubsection{Intrinsic wave functions for $^{20}$Ne($0^+$) and $^{20}$Ne($1^-$)}
For $^{20}$Ne($0^+_1$) and $^{20}$Ne($1^-_1$), I use $^{16}{\rm O}+\alpha$-cluster 
wave functions.
Zhou {\it et al.} have shown that these states of $^{20}$Ne are well described by
3D-THSR wave functions of $^{16}{\rm O}+\alpha$ \cite{Zhou:2012zz,Zhou:2013ala}.
In the present paper, I consider the spherical 3D-THSR wave functions, which can have more than 98\% and 99\%
overlaps with the exact solutions of the $^{16}{\rm O}+\alpha$ states for $^{20}$Ne($0^+$)
and $^{20}$Ne($1^-$) \cite{Kanada-En'yo:2014vva}.

In the present paper, I fix the center position of 
$^{16}$O at the origin by omitting the recoil of the $^{16}$O core 
and consider the $R^k$-weighted Gaussian distribution 
of an $\alpha$ cluster around the $^{16}$O core. 
The intrinsic wave function of the 3D-THSR wave function of
$^{16}{\rm O}+\alpha$ is essentially given by the following separable form of
the 1D-THSR wave function of $\alpha$-($2\alpha$) on the $z$ axis 
and the $(0p_x)^4 (0p_y)^4$ ho shell-model wave function as
\begin{eqnarray}
\Phi^{\rm int}_{^{20}{\rm Ne}}(R^k;\beta)&=&
{\cal A}\left\{\Phi_{\rm ho}\left[(0p_x)^4 (0p_y)^4 \right]
\Phi^{\alpha\textrm{-}(2\alpha)}_\textrm{1D-THSR}(R^k;\beta)\right\}\\
\Phi^{\alpha\textrm{-}(2\alpha)}_\textrm{1D-THSR}(R^k;\beta)&=&\int dR_1 R_1^k
\exp\left\{ -\frac{R_1^2}{\beta^2}
\right\}\Phi^{3\alpha}_{\rm BB}(R_1,R_2=+\varepsilon,R_3=-\varepsilon), \label{eq:2a-a}
\end{eqnarray}
where $\Phi_{\rm ho}\left[(0p_x)^4 (0p_y)^4 \right]$ is the 8-nucleon
wave function of the $(0p_x)^4 (0p_y)^4$ ho shell-model configuration at the origin, and 
$\Phi^{\alpha\textrm{-}(2\alpha)}_\textrm{1D-THSR}(R^k;\beta)$ is the 1D-THSR 
${\alpha\textrm{-}(2\alpha)}$ wave function for an $\alpha$ cluster 
with the $R^k$-weighted Gaussian distribution around the fixed $2\alpha$ core at the origin,
and $\varepsilon$ is taken to be an enough small value.
$\Phi^{\rm int}_{^{20}{\rm Ne}}(R^k;\beta)$ is regarded as the intrinsic
wave function of the spherical 3D-THSR wave function
for $^{20}$Ne($0^+$) and $^{20}$Ne($1^-$)  in the no recoil approximation
as 
\begin{eqnarray}
P^{J=0}\Phi^{\rm int}_{^{20}{\rm Ne}}(R^2;\beta)&\approx&
\int d\bvec{R}_1 
\exp\left\{ -\frac{\bvec{R}_1^2}{\beta^2}\right\}
\Phi^{^{16}{\rm O}\textrm{-}\alpha}_{\rm BB}
(\bvec{R}'=0, \bvec{R}_1),\\
P^{J=1}_M\Phi^{\rm int}_{^{20}{\rm Ne}}(R^3;\beta)&\approx&
\int d\bvec{R}_1 
R_1 Y_{1M}(\hat{\bvec{R}}_1)\exp\left\{ -\frac{\bvec{R}_1^2}{\beta^2}\right\}
\Phi^{^{16}{\rm O}\textrm{-}\alpha}_{\rm BB}
(\bvec{R}'=0, \bvec{R}_1),\\
\Phi^{^{16}{\rm O}\textrm{-}\alpha}_{\rm BB}
(\bvec{R}', \bvec{R}_1)&=& 
{\cal A}\left[
\psi^{^{16}{\rm O}}_{\bvec{R}'}
\psi^\alpha_{\bvec{R}_1} \right].
\end{eqnarray}
Here $\psi^{^{16}{\rm O}}_{\bvec{R}'}$ is 
the ho shell-model wave function of the $p$-shell closure around the position 
$\bvec{R}'$. 

In the intrinsic wave function $\Phi^{\rm int}_{^{20}{\rm Ne}}(R^k;\beta)$, 
8 nucleons in $0p_x$ and $0p_y$ orbits give no contribution to the 
entanglement entropy,  and therefore, I analyze
the 1D-THSR wave function of $\alpha\textrm{-}(2\alpha)$, 
$\Phi^{\alpha\textrm{-}(2\alpha)}_\textrm{1D-THSR}(R^k;\beta)$, for 
$^{20}$Ne.
In the practical calculation, the $R_1$ integration in Eq.~\eqref{eq:2a-a} is approximated by 
summation of mesh points, 
$\sum_{R=0,\pm 1,\ldots,\pm 12\ {\rm fm}}$, and 
$\varepsilon=0.02$ fm is used. 

\section{Results}\label{sec:results}
I analyze the system size ($\beta$) dependence of the entanglement entropy in 1D 
cluster states of 
$n\alpha$ and $^{16}{\rm O}+\alpha$ systems. Based on the analysis of the entanglement entropy, 
I discuss the delocalization of clusters in the linear-chain $3\alpha$ and $4\alpha$
states, the $2\alpha$ state for $^8$Be, and the $^{16}{\rm O}+\alpha$ states for $^{20}$Ne, 
whose intrinsic wave functions are approximately described by the 1D-THSR wave functions, 
$\Phi^{n\alpha}_{\textrm{1D-THSR}}(\beta)$, $\Phi^{\rm int}_{^8{\rm Be}}(R^2;\beta)$, and
$\Phi^{\alpha\textrm{-}(2\alpha)}_{\textrm{1D-THSR}}(R^k;\beta)$, respectively, with optimum $\beta$ values.
The optimum $\beta$ values for the linear-chain $3\alpha$ and $4\alpha$
states are taken from Ref.~\cite{Suhara:2013csa}, and those for $^8$Be($0^+_1$),
$^{20}$Ne($0^+$), and $^{20}$Ne($1^-$) are reduced from the results for the 
spherical 3D-THSR wave functions in Ref.~\cite{Kanada-En'yo:2014vva}.

\subsection{Analysis of one cluster}

\begin{figure}[htb]
\begin{center}
	\includegraphics[width=5.5cm]{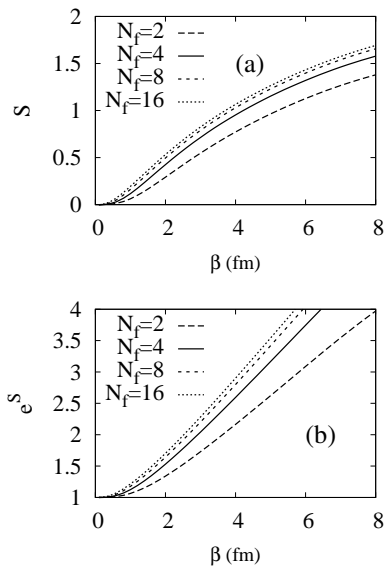} 	
\end{center}
  \caption{(a) $\beta$ dependence of entanglement entropy 
$S$ for a cluster consisting of $N_f$ particles for $N_f=2$, 4, 8, and 16. 
$S$ defined by the reduced density matrix for a fixed species $\sigma$ is shown.
(b) Same as (a) but $\beta$ dependence of $\exp(S)$.
\label{fig:s-mp}}
\end{figure}

I analyze entanglement entropy of a $1\alpha$ system expressed by
the 1D-THSR wave function $\Phi^{1\alpha}_\textrm{1D-THSR}(\beta)$, where
an $\alpha$ cluster moves in the Gaussian distribution with the range $\beta$ 
around the origin. 
The c.m.m. is not corrected. This system is regarded 
as an $\alpha$ cluster trapped in a ho external potential. 
In $\Phi^{1\alpha}_\textrm{1D-THSR}(\beta)$,  $\alpha$ is the composite
particle of 4 nucleons, but it is easy to mathematically extend the 1D-THSR wave function for 
a general composite particle consisting of $N_f$ 
constituent particles ($N_f$ is the particle number in the composite particle), 
\begin{equation}
\psi^{\alpha}_{\bvec{R}_1}=\prod_{\sigma=1}^{N_f}
\phi^{\rm 0s}_{\bvec{R}_1}\chi_\sigma,
\end{equation}
where $\sigma$ is the label for $N_f$ species of particles, which are not identical to each other. 
Figure \ref{fig:s-mp}(a) shows $\beta$ dependence of the entanglement entropy 
$S$ defined by the reduced density matrix for a $\sigma$ particle. 
In the small system size ($\beta$) limit, 
$S$ goes to zero because the cluster is localized around the origin in the 1D-THSR wave function. 
With  increase of the system size $\beta$, the entanglement entropy $S$ increases. 
For a fixed $\beta$, $S$ is saturated with increase of  the number of constituent particles ($N_f$).

If the one-body density is fragmented equally into $m$ states, $S=\log m$ 
($m$ is the number of states). 
For instance, in the case that the one-body density is fragmented equally into two states, 
$S=\log 2=0.693$.
It is naively expected that the number of independent states that contribute to the entanglement entropy
$S$ is proportional to $\beta/b$ (the system size divided by the cluster size).
Provided that the one-body density is fragmented equally into these states, 
$\exp(S)$ should have a linear dependence 
on the system size $\beta$.  As shown 
in Fig.~\ref{fig:s-mp}(b), $\exp(S)$ is a almost linear function of $\beta$, 
supporting the naive expectation.

\begin{figure}[htb]
\begin{center}
	\includegraphics[width=11cm]{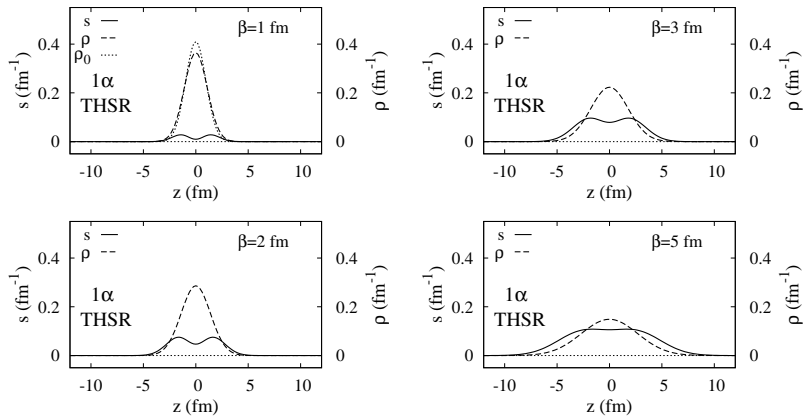} 	
\end{center}
  \caption{Spatial distributions of 
local entanglement entropy $s(z)$ and density $\rho(z)$ 
in the 1D-THSR wave function for the $1\alpha$ system with 
$\beta=1$, 2, 3, and 5 fm. The density $\rho_0(z)$ of a 
localized $\alpha$ for the $\beta=0$ limit is also shown in the left top panel.
\label{fig:rho-1a}}
\end{figure}

Spatial distributions of local entanglement entropy $s(z)$ and density $\rho(z)$ in
the 1D-THSR wave function for the $1\alpha$ system is shown 
in Fig.~\ref{fig:rho-1a}.
In case of the system size as small as $\beta=1$ fm, where an $\alpha$ cluster is well localized 
around the origin as shown by the density almost the same as that for a fixed $\alpha$ cluster, 
$s(z)$ is quite small. As $\beta$ increases, delocalization 
of cluster develops and $s(z)$ increases 
in particular in low-density regions at the surface.
$s(z)$ is relatively suppressed at the high-density region near the origin. 

The analysis of one-cluster systems suggests that the entanglement entropy is enhanced  
in a low-density cluster gas having a relatively larger system size than the cluster size.
 The entanglement entropy $S$ and the local entropy 
$s(z)$ can be a good measure for the delocalization of cluster. 

\subsection{Linear-chain states of $n\alpha$ systems}

\begin{figure}[htb]
\begin{center}
	\includegraphics[width=5.5cm]{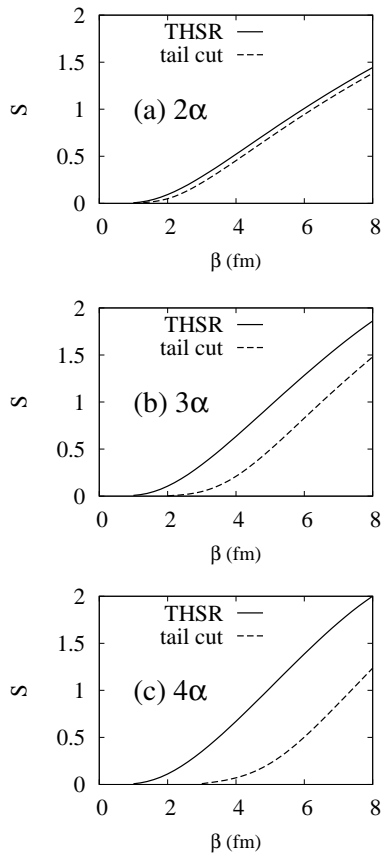} 	
\end{center}
  \caption{$\beta$ dependence of entanglement entropy 
$S$ of the 1D-THSR wave functions 
for $2\alpha$, $3\alpha$, and $4\alpha$ (solid lines). $S$ of the tail-cut 
1D-THSR wave functions is also shown (dashed lines). 
\label{fig:s-2a-3a-4a}}
\end{figure}

\begin{figure}[htb]
\begin{center}
	\includegraphics[width=5.5cm]{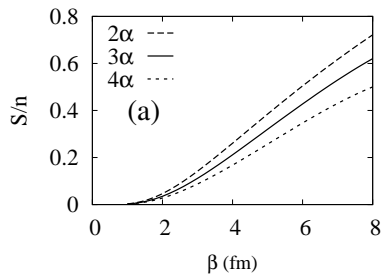} 	
\end{center}
  \caption{Entanglement entropy 
per $\alpha$ cluster, $S/n$, plotted as a function of $\beta$ 
of the 1D-THSR wave functions 
for $2\alpha$, $3\alpha$, and $4\alpha$.
\label{fig:s-normalize}}
\end{figure}

I analyze entanglement entropy of the 1D-THSR wave functions 
for the linear-chain $2\alpha$, $3\alpha$, and $4\alpha$ states. 
Figure \ref{fig:s-2a-3a-4a} shows the system size ($\beta$) dependence 
of $S$. Similarly to the $1\alpha$ system discussed previously, 
$S$ increases as the system size $\beta$ increases.
In Fig.~\ref{fig:s-normalize}, 
entanglement entropy 
per $\alpha$ cluster, $S/n$,  is plotted as a function of $\beta$.
If  $\alpha$-cluster motion is not affected by other $\alpha$ clusters,
$S$ should be proportional to the number ($n$) of $\alpha$ clusters.
$S/n$ is somewhat suppressed as $n$ increases because of the Pauli blocking effect 
between $\alpha$ clusters, which comes from 
the antisymmetrization effect between nucleons in different $\alpha$ clusters.
That is, delocalization of $\alpha$ clusters is suppressed because 
the effective system size for the $\alpha$-cluster motion is reduced by 
the Pauli blocking effect. 

The Pauli blocking
effect should be relatively strong in the inner high-density region 
rather than at the low-density surface region.
To see the contribution of the low-density tail part of the $\alpha$-cluster Gaussian distribution 
to the entanglement entropy, 
I calculate $S$ of the ``tail-cut'' 1D-THSR wave functions with 
no outer tail, where the basis wave functions are truncated as $|R'_i| \le \beta$
by cutting off the basis wave functions in the $|R'_i| > \beta$ region for the tail component. 
$S$ of the tail-cut 1D-THSR wave functions 
indicates the contribution of the inner part in the total entropy,  whereas, 
the difference of $S$ between the tail-cut wave functions and the original ones approximately 
corresponds to the contribution of the tail part.
$S$ of the tail-cut 1D-THSR wave functions shown by dashed lines in Fig.~\ref{fig:s-2a-3a-4a} 
is much smaller  than that of the original 1D-THSR wave functions in $3\alpha$ and $4\alpha$ systems,
indicating the significant contribution of the tail component in $S$. 
The tail contribution in the total entanglement entropy is larger than 50\%
in the $\beta\le 4$ fm region for $3\alpha$ and 
in the $\beta\le 6$ fm region for $4\alpha$. For the same $\beta$ value, the contribution of the inner part in $S$ 
is relatively small in $4\alpha$ than in 3$\alpha$ because of the Pauli blocking effect.

\begin{figure}[htb]
\begin{center}
	\includegraphics[width=16.5cm]{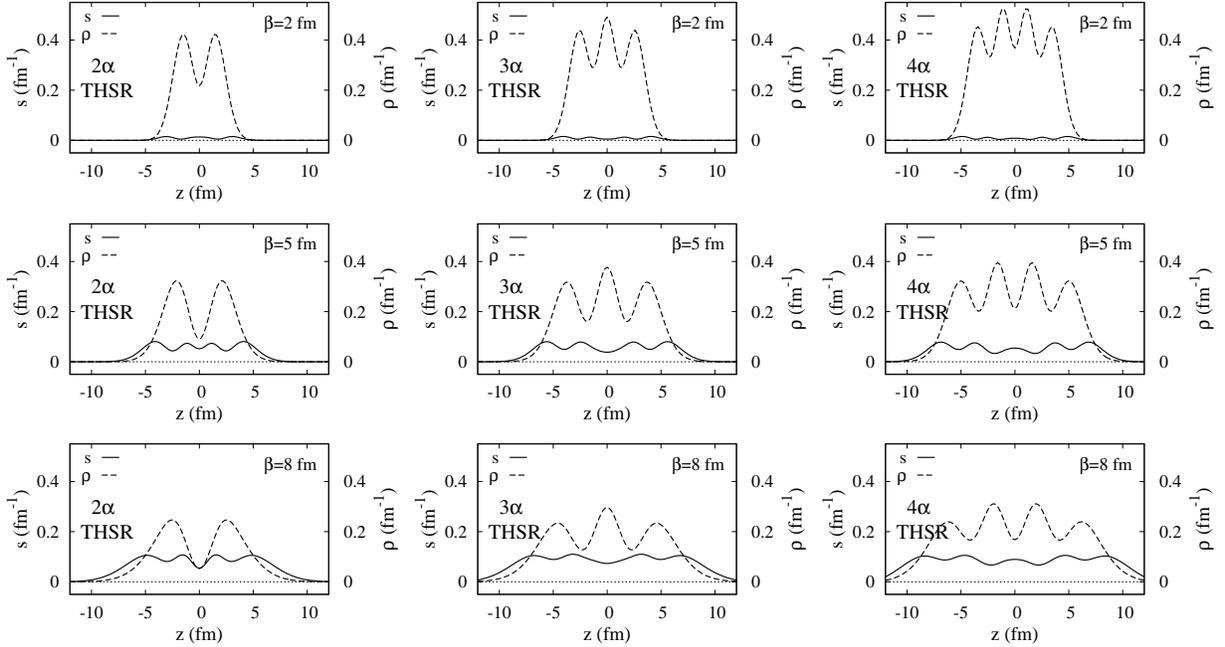} 	
\end{center}
  \caption{Spatial distributions of 
local entanglement entropy $s(z)$ and density $\rho(z)$ 
in the 1D-THSR wave functions of $2\alpha$, $3\alpha$, and $4\alpha$ systems
with $\beta=2$, 5, and 8 fm.
\label{fig:rho-2a-3a-4a}}
\end{figure}

Figure \ref{fig:rho-2a-3a-4a} shows local entanglement entropy $s(z)$ of $2\alpha$, $3\alpha$, and 
$4\alpha$ for the system size $\beta=2$, 5, and 8 fm. In case of the small system size as $\beta=2$ fm, 
$s(z)$ is almost zero, which indicates no delocalization of $\alpha$ clusters. 
As $\beta$ increases, $s(z)$ increases and shows a broader spatial distribution 
than the density distribution. $s(z)$ is relatively suppressed in the inner high-density 
region and enhanced in the low-density region at the surface.
It means that the significant contribution in the entanglement entropy 
originates in more broadly distributed orbits than the system size.
Comparing the result for a given $\beta$ between $2\alpha$, $3\alpha$, and $4\alpha$ systems, 
the larger $n$ (the number of $\alpha$ clusters) system 
shows more suppression of $s(z)$ in the inner region  than the smaller $n$ system
because of the Pauli blocking effect.

\begin{figure}[htb]
\begin{center}
	\includegraphics[width=11cm]{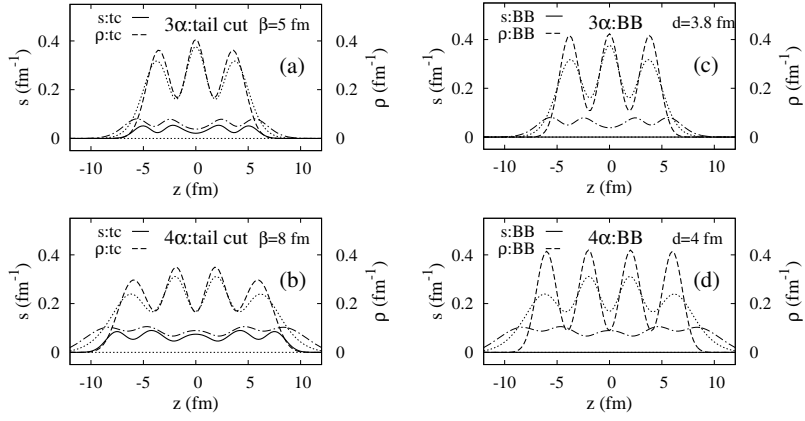} 	
\end{center}
  \caption{Spatial distributions of 
local entanglement entropy $s(z)$ and density $\rho(z)$ 
in (a) the tail-cut 1D-THSR wave function of $3\alpha$ with $\beta=5$ fm, (b)
the tail-cut 1D-THSR wave function of $4\alpha$ with $\beta=8$ fm, 
(c) the BB wave function of $3\alpha$ with the interval $d=3.8$ fm, 
and (d) the BB wave function of $4\alpha$ with $d=4$ fm. 
$s(z)$ and $\rho(z)$ in the original 1D-THSR wave functions of $3\alpha$ with $\beta=5$ fm 
are also shown by dash-dotted and dotted lines, respectively, in  (a) and (c), 
and those of $4\alpha$ with $\beta=8$ fm are shown in (b) and (d).
\label{fig:rho-3a-4a-BB}}
\end{figure}

Let us consider correspondence of the present result with 
the linear-chain $3\alpha$ and $4\alpha$ states in $^{12}$C and $^{16}$O
predicted by Suhara {\it et al.}
In Ref.~\cite{Suhara:2013csa}, they have applied 
the generator coordinate method (GCM)
to the linear-chain 
$3\alpha$ and $4\alpha$ states using effective nuclear forces to exactly solve 1D dynamics 
of $3\alpha$ and $4\alpha$,
and obtained the energy minimum wave functions in the model space of linear configurations. 
It has been shown that the
wave functions of the linear $3\alpha$ and $4\alpha$ states 
obtained with the GCM calculations
have large overlap with the 1D-THSR wave functions.
The optimum parameters of the 1D-THSR wave functions
are $\beta=5.1$ fm and $\beta=8.2$ fm for the $0^+$ linear-chain states 
of $3\alpha$ and $4\alpha$, respectively. Looking at the result of corresponding wave functions, 
$\Phi^{3\alpha}_{\textrm{1D-THSR}}(\beta=5$ fm) and $\Phi^{4\alpha}_{\textrm{1D-THSR}}(\beta=8$ fm), 
shown in Figs.~\ref{fig:s-2a-3a-4a} and \ref{fig:rho-2a-3a-4a}, it is found that the 
entanglement entropy is generated by the delocalization of $\alpha$ clusters. In particular, 
in the linear-chain $4\alpha$ states with the large system size as $\beta=8$ fm, $s(z)$ is broadly distributed with 
significant amplitude.
In both cases of the linear-chain $3\alpha$ and $4\alpha$ states, 
the tail contribution is significantly large as shown in the comparison of 
$S$ between the tail-cut 1D-THSR and the original 1D-THSR wave functions
in Fig.~\ref{fig:s-2a-3a-4a}. 
In the $3\alpha$ state
with $\beta=5$ fm, the tail contribution is as large as $\sim 50\%$ of the total entropy.
In Fig.~\ref{fig:rho-3a-4a-BB},
spatial distributions of $s(z)$ and $\rho(z)$ in the tail-cut 
1D-THSR wave function of $3\alpha$ with $\beta=5$ fm and those of 
$4\alpha$ with $\beta=8$ fm are shown by solid and dashed lines
compared with those in the original 1D-THSR
wave functions shown by dash-dotted and dotted lines.
For the $3\alpha$ state with $\beta=5$ fm, 
the tail-cut 1D-THSR wave function shows the overall reduction of $s(z)$
compared with the original 1D-THSR, indicating that 
the delocalization of $\alpha$ clusters occurs mainly in the low-density region at the surface.
In other words, about half of the 
total entanglement entropy is generated due to the small difference 
in the tail component of the $\alpha$ distribution 
of the original 1D-THSR wave function from the tail-cut 1D-THSR one.
In the $4\alpha$ state with $\beta=8$ fm,  
the tail-cut 1D-THSR wave function shows some reduction of $s(z)$
at the surface region, but a significant amplitude of $s(z)$ still remains 
in the inner region even after the tail cut. It means that, in the $4\alpha$ state, 
the delocalization of $\alpha$ clusters
occurs also in the inner region as well as at the surface region.

I also demonstrate $s(z)$ and $\rho(z)$ in the localized cluster limit of the
linear-chain $3\alpha$ and $4\alpha$ states given by the BB wave functions, 
where each $\alpha$ cluster is localized at a certain position $R_i$.
They corresponds to the conventional linear-chain states.
I set the $\alpha$-cluster positions $R_i$ on the $z$ axis at equal intervals 
$d=3.8$ fm in the $3\alpha$ system and 
$d=4$ fm in the $4\alpha$ system so as to give density ($\rho(z)$) peak positions similar to those of
the corresponding 1D-THSR wave functions, 
$\Phi^{3\alpha}_{\textrm{1D-THSR}}(\beta=5$ fm) and $\Phi^{4\alpha}_{\textrm{1D-THSR}}(\beta=8$ fm).
In Fig.~\ref{fig:rho-3a-4a-BB},
$s(z)$ and $\rho(z)$ in these BB wave functions for the conventional linear-chain states
are compared with those of the 1D-THSR wave functions.
$s(z)$ in the BB wave functions completely vanishes 
because $S$ is trivially zero for a single Slater determinant. 
This result indicates that 
local entanglement entropy $s(z)$ shows 
the prominent difference between the localized and delocalized wave functions 
for the linear-chain $3\alpha$ and $4\alpha$ states 
even though the difference in density 
is not so remarkable, in particular, in the 3$\alpha$ state.
It is concluded that both $S$ and $s(z)$ are sensitive to the delocalization 
of $\alpha$ clusters.

\subsection{$\alpha$+$\alpha$ for $^{8}$Be}

As pointed out by Funaki {\it et al.},
the exact $2\alpha$ wave function for 
$^8$Be($0^+_1$) is well described  by the 3D-THSR wave function  \cite{Funaki:2002fn,Funaki:2009zz}.
In the present paper, I use the $R^2$-weighted 1D-THSR wave function, $\Phi^{\rm int}_{^8{\rm Be}}(\beta)$, in Eq.~\eqref{eq:int-8Be}
as the intrinsic wave function of the spherical 3D-THSR wave function.
The parameter $\beta=3.3$ fm is reduced from 
the optimum parameter $B=4.77$ fm with $b=1.36$ fm  of the 
spherical 3D-THSR wave function for $^8$Be($0^+_1$) taken from
Ref.~\cite{Kanada-En'yo:2014vva},
using the relation $B^2=b^2+2\beta^2$ and the scaling 
$b=1.36\ {\rm fm}\rightarrow 1.376$ fm. 
 (The parameter $B$ here is originally 
labeled by ``$\sigma$" in Ref.~\cite{Kanada-En'yo:2014vva}.) 

\begin{figure}[htb]
\begin{center}
	\includegraphics[width=5.5cm]{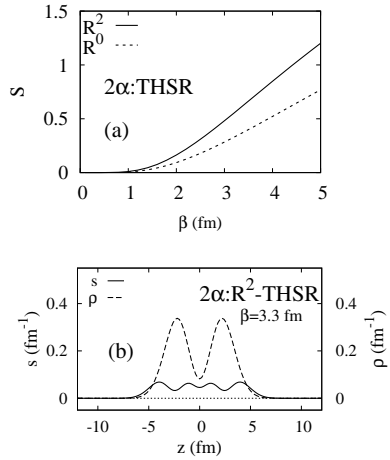} 	
\end{center}
  \caption{(a) $\beta$ dependence of entanglement entropy 
$S$ of the $R^2$-weighted 1D-THSR $\Phi^{\rm int}_{^8{\rm Be}}(\beta)$ of 
$2\alpha$.  That of the normal ($R^0$-weighted) 
1D-THSR of $2\alpha$ is also shown. 
(b) Spatial distributions of 
local entanglement entropy $s(z)$ and density $\rho(z)$ 
in $\Phi^{\rm int}_{^8{\rm Be}}(\beta)$ with $\beta=3.3$ fm
for $^8$Be$(0^+_1)$. 
\label{fig:2a.l0}}
\end{figure}

Figure \ref{fig:2a.l0}(a) shows $\beta$ dependence of the entanglement entropy $S$
of  $\Phi^{\rm int}_{^8{\rm Be}}(\beta)$. For comparison, 
$S$ of the $R^0$-weighted 1D-THSR wave function is also shown. 
At the optimized parameter $\beta=3.3$ fm for $^8$Be($0^+_1$), 
$S=0.6$ is generated by the delocalization of $\alpha$ clusters.
Spatial distributions of local entanglement entropy 
($s(z)$) and density ($\rho(z)$) in $\Phi^{\rm int}_{^8{\rm Be}}(\beta=3.3 \ {\rm fm})$ are shown in Fig.~\ref{fig:2a.l0}(b).

\subsection{$^{16}$O+$\alpha$ for $^{20}$Ne}

As shown by Zhou {\it et al.}, $^{20}$Ne($0^+_1$) and $^{20}$Ne($1^-_1$) 
are well described by the 3D-THSR wave functions \cite{Zhou:2012zz,Zhou:2013ala}. 
As approximated intrinsic wave functions of the 
spherical 3D-THSR wave functions for $^{20}$Ne($0^+_1$) and $^{20}$Ne($1^-_1$),
I use the $R^2$- and $R^3$-weighted 1D-THSR 
wave functions, $\Phi^{\alpha\textrm{-}(2\alpha)}_\textrm{1D-THSR}(R^k;\beta)$ in Eq.~\eqref{eq:2a-a}, respectively.
For comparison, 
I also use localized cluster wave functions given by
following parity projected BB wave functions with the fixed 
$\alpha$-$(2\alpha)$ distance $d$, 
\begin{eqnarray}
\Phi^{\alpha\textrm{-}(2\alpha),\pm}_{\rm BB}(d)&=&
\Phi^{3\alpha}_{\rm BB}(R_1=d,R_2=+\varepsilon,R_3=-\varepsilon)\nonumber\\
&&\pm\Phi^{3\alpha}_{\rm BB}(R_1=-d,R_2=+\varepsilon,R_3=-\varepsilon).
\end{eqnarray}
$\Phi^{\alpha\textrm{-}(2\alpha),\pm}_{\rm BB}(d)$ is given by the linear combination of 
two Slater determinants, and in the large $d$ limit, it has $S=\log 2$. 
For the 1D-THSR $\alpha$-$(2\alpha)$ wave functions, 
the optimum parameters $\beta=2.0$ fm and $\beta=2.6$ fm for $^{20}$Ne($0^+_1$) and $^{20}$Ne($1^-_1$)  
are reduced from $B=2.39$ fm and  $B=2.97$ fm, respectively for the spherical 3D-THSR 
wave functions with $b=1.46$ fm in Ref.~\cite{Kanada-En'yo:2014vva} 
using the relation $B^2=\beta^2+b^2/2$ and the scaling $b=1.46\ {\rm fm}\rightarrow 1.376$ fm.
For the positive- and negative-parity projected BB wave functions, 
I use the optimum parameters $d=3.05$ fm and $d=3.85$ fm
taken from Ref.~\cite{Kanada-En'yo:2014vva}, respectively.

\begin{figure}[htb]
\begin{center}
	\includegraphics[width=5.5cm]{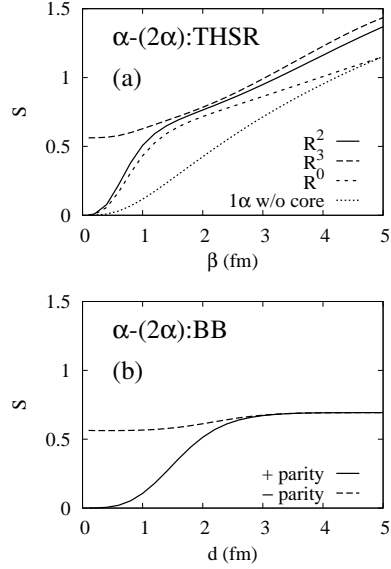} 	
\end{center}
  \caption{(a) $\beta$ dependence of entanglement entropy 
$S$ of the $R^k$-weighted 1D-THSR wave functions 
$\Phi^{\alpha\textrm{-}(2\alpha)}_\textrm{1D-THSR}(R^k;\beta)$ of $\alpha$-$(2\alpha)$
and the 1D-THSR wave function $\Phi^{1\alpha}_\textrm{1D-THSR}(\beta)$
of $1\alpha$ without the $2\alpha$ core, 
and (b) $d$ dependence of $S$ of the parity projected BB wave functions
$\Phi^{\alpha\textrm{-}(2\alpha),\pm}_{\rm BB}(d)$.
\label{fig:s-20Ne-beta}}
\end{figure}

Fig.~\ref{fig:s-20Ne-beta} shows 
$\beta$ dependence of $S$ 
of the 1D-THSR wave function, 
$\Phi^{\alpha\textrm{-}(2\alpha)}_\textrm{1D-THSR}(R^k;\beta)$ for $^{20}$Ne, 
and also shows $d$ dependence of $S$ for the BB wave function,
$\Phi^{\alpha\textrm{-}(2\alpha),\pm}_{\rm BB}(d)$ for the localized cluster wave function. 
For comparison, I also show the result of the 
$R^0$-weighted 1D-THSR wave function of an $\alpha$ cluster 
with and without the $2\alpha$ core.
In the $R^2$-weighted 1D-THSR wave function of $\alpha$-$(2\alpha)$, 
$S$ is zero at $\beta\rightarrow 0$ and rapidly increases in $\beta \lesssim 1$, and it gradually increases 
in the $\beta \gtrsim 1$ region.
In the $R^3$-weighted 1D-THSR wave function, $S$ is finite even at $\beta\rightarrow 0$ 
because $\Phi^{\alpha\textrm{-}(2\alpha)}_\textrm{1D-THSR}(R^3;\beta\rightarrow 0)$
is equivalent to the shell model limit wave function 
having 3 nucleons in the $sd$ shell and a nucleon in the $pf$ shell, which has $S=-\frac{3}{4}\log\frac{3}{4}-
\frac{1}{4}\log \frac{1}{4} =0.562$.
With increase of $\beta$, $S$ of $\Phi^{\alpha\textrm{-}(2\alpha)}_\textrm{1D-THSR}(R^3;\beta)$ becomes close to that of $\Phi^{\alpha\textrm{-}(2\alpha)}_\textrm{1D-THSR}(R^2;\beta)$. 
In case of the positive-parity projected BB wave function, 
$S$ rapidly increases with the increase of $d$ and it becomes constant $S=\log 2$ in the $d\gtrsim 3$ fm region. 
The negative-parity projected BB wave function shows small $d$ dependence of $S$ as 
$S=0.562$ at $d\rightarrow 0$ and $S=\log 2$ in the large $d$ region.
$S=\log 2$ is generated in both the positive- and negative-parity projected BB wave functions with a large $d$, 
because the $\alpha$-cluster wave functions are separated into two parts in 
$z>0$ and $z<0$, which have almost no overlap with each other. 
This indicates that even the localized cluster wave functions can have the finite
entanglement entropy $S=\log 2$ by the parity projection.
In comparison of $S$ between the 1D-THSR and BB wave functions, 
it is found that 
the entanglement entropy $S\sim \log 2$  generated in 
the 1D-THSR wave functions of $\alpha$-$(2\alpha)$ with $\beta\lesssim 1.5$ fm does not originate 
in the delocalization of the $\alpha$ cluster but it is understood by the effect of parity projection.
With further increase of $\beta$ in the 1D-THSR wave functions, $S$ increases due to the delocalization of 
the $\alpha$ cluster and it becomes larger than $\log 2$. 

\begin{figure}[htb]
\begin{center}
	\includegraphics[width=11cm]{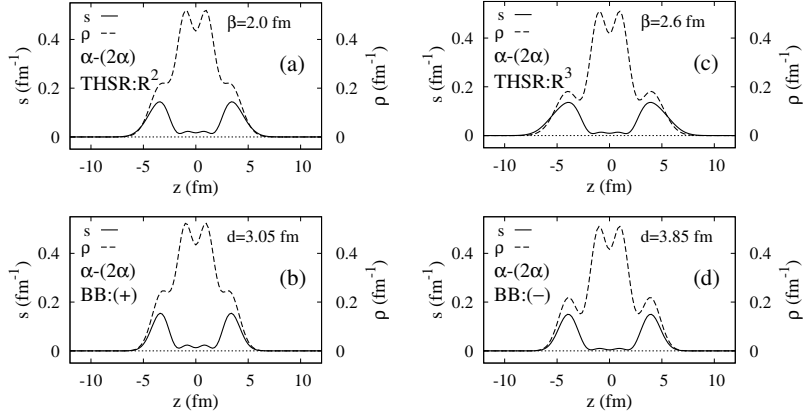} 	
\end{center}
  \caption{Spatial distributions of 
local entanglement entropy $s(z)$ and density $\rho(z)$ 
in (a) $\Phi^{\alpha\textrm{-}(2\alpha)}_\textrm{1D-THSR}(R^2;\beta)$ with
$\beta=2.0$ fm for $^{20}$Ne($0^+_1$), 
(b) $\Phi^{\alpha\textrm{-}(2\alpha),+}_{\rm BB}(d)$ with $d=3.05$ fm, 
(c) $\Phi^{\alpha\textrm{-}(2\alpha)}_\textrm{1D-THSR}(R^3;\beta)$ with
$\beta=2.6$ fm for  $^{20}$Ne($1^-_1$), and (d) $\Phi^{\alpha\textrm{-}(2\alpha),-}_{\rm BB}(d)$ with $d=3.85$ fm.
\label{fig:rho-2a-a}}
\end{figure}

Let us consider correspondence of the present result with 
$^{20}$Ne($0^+_1$) and $^{20}$Ne($1^-_1$). As mentioned previously, 
$\Phi^{\alpha\textrm{-}(2\alpha)}_\textrm{1D-THSR}(R^2;\beta=2.0\ {\rm fm})$
and $\Phi^{\alpha\textrm{-}(2\alpha)}_\textrm{1D-THSR}(R^3;\beta=2.6\ {\rm fm})$
correspond to the intrinsic wave functions of 
$^{20}$Ne($0^+_1$) and $^{20}$Ne($1^-_1$) described by the spherical 3D-THSR wave functions, respectively. 
$\Phi^{\alpha\textrm{-}(2\alpha),+}_{\rm BB}(d=3.05\ {\rm fm})$ and 
 $\Phi^{\alpha\textrm{-}(2\alpha),-}_{\rm BB}(d=3.85\ {\rm fm})$  
correspond to the optimized BB wave functions for $^{20}$Ne($0^+_1$) and $^{20}$Ne($1^-_1$).
Figure \ref{fig:rho-2a-a} shows spatial distributions of 
local entanglement entropy ($s(z)$) and density ($\rho(z)$) in these wave functions. 
For $^{20}$Ne($0^+_1$), $\Phi^{\alpha\textrm{-}(2\alpha)}_\textrm{1D-THSR}(R^2;\beta=2.0 \ {\rm fm})$ and $\Phi^{\alpha\textrm{-}(2\alpha),+}_{\rm BB}(d=3.05\ {\rm fm})$ show 
similar distributions of $s(z)$ and $\rho(z)$ to each other. 
 $s(z)$ almost vanishes in the $-2\le z\le2$ fm region because of the Pauli blocking from the 
core and it has amplitude only at surface regions. 
$\Phi^{\alpha\textrm{-}(2\alpha)}_\textrm{1D-THSR}(R^2;\beta=2.0\ {\rm fm})$ and $\Phi^{\alpha\textrm{-}(2\alpha),+}_{\rm BB}(d=3.05 \ {\rm fm})$ have 
the total entanglement entropy $S=0.76$ and $S=0.67$, respectively. 
They are close to $\log 2$, which is generated mainly just by the parity projection, 
indicating that the delocalization of the $\alpha$ cluster is weak
in the intrinsic state of $^{20}$Ne($0^+_1$). 
For $^{20}$Ne($1^-_1$), $\Phi^{\alpha\textrm{-}(2\alpha)}_\textrm{1D-THSR}(R^3;\beta=2.6\ {\rm fm})$ shows a broader distribution of $s(z)$ in the outer tail part than that of $\Phi^{\alpha\textrm{-}(2\alpha),-}_{\rm BB}(d=3.85\ {\rm fm})$. 
$\Phi^{\alpha\textrm{-}(2\alpha)}_\textrm{1D-THSR}(R^3;\beta=2.6\ {\rm fm})$ has the total entropy
$S=0.90$ larger than $S=0.69$ of  
$\Phi^{\alpha\textrm{-}(2\alpha),-}_{\rm BB}(d=3.85\ {\rm fm})$. 
It means that, the entropy is somewhat generated by the delocalization of the $\alpha$ cluster 
in addition to $S=\log 2$ from the parity projection.
This additional entanglement entropy by the delocalization comes from
a low-density outer tail of the wave function. That is, the delocalization of an $\alpha$ cluster
occurs in the low-density tail part.

\section{Summary and outlook}\label{sec:summary}
In order to investigate delocalization of clusters in the 1D cluster systems, 
I proposed a method of analysis using entanglement entropy defined by the one-body 
density matrix. 
I studied  the entanglement entropy of the 
1D cluster states of $n\alpha$ and $^{16}{\rm O}+\alpha$ systems,
and discussed the delocalization of clusters in the intrinsic wave functions of the linear-chain $3\alpha$- and $4\alpha$-cluster states, $^8$Be($0^+_1$), and 
$^{20}$Ne($0^+_1,1^-_1$).
I investigated the entanglement entropy of the 1D-THSR wave functions and compared it with that of 
the BB cluster wave functions, and showed clear differences in the entanglement entropy 
between localized cluster wave functions and delocalized cluster wave functions.

I calculated the entanglement entropy of the 1D-THSR wave functions for 
linear-chain $2\alpha$, $3\alpha$, and $4\alpha$ systems, and discuss the dependences on the 
system size $(\beta)$ and the number ($n$) of $\alpha$ clusters. With increase of the system size $\beta$, 
the entanglement entropy increases as the delocalization of $\alpha$ clusters develops.
With increase of the number of $\alpha$ clusters, the entanglement entropy per $\alpha$ cluster decreases because of the 
Pauli blocking effect between clusters. 
In order to clarify the spatial regions where the entanglement entropy is generated by the delocalization, 
I defined the local entanglement entropy $s(z)$. I found that 
the entanglement entropy is generated in the low-density part whereas it is relatively suppressed in the high-density part, 
indicating that the delocalization of clusters occurs dominantly in the low-density region but
it is suppressed in the high-density region because of the Pauli blocking effect between clusters. 

Moreover, I discussed the delocalization of $\alpha$ clusters in 
the linear-chain $3\alpha$ and $4\alpha$ states  predicted in $^{12}$C and $^{16}$O by Suhara {\it et al.} \cite{Suhara:2013csa}
based on the analysis of the entanglement entropy.
In the linear-chain $3\alpha$ state, the delocalization of clusters occurs dominantly in the low-density tail region while
it is relatively suppressed in the inner region because of the Pauli blocking effect. 
In the linear-chain 4$\alpha$ state having the significantly larger system size 
than the linear-chain $3\alpha$ state, 
the delocalization occurs in the whole system. 

I also analyzed the entanglement entropy of the $R^k$-weighted 1D-THSR wave functions of
$2\alpha$ and $\alpha$-$(2\alpha)$ systems, which correspond to the intrinsic wave functions of 
the cluster states in $^8$Be($0^+_1$), $^{20}$Ne$(0^+_1)$, and $^{20}$Ne$(1^-_1)$. 
In $^8$Be($0^+_1$),  
the entanglement entropy is generated because of the delocalization of clusters.
In $^{20}$Ne$(0^+_1)$ and $^{20}$Ne$(1^-_1)$, the entanglement entropy is strongly suppressed in the inner region because of the
Pauli blocking from the core. 
The entanglement entropy is generated at the surface region mainly because of the parity projection. 
In particular, in $^{20}$Ne$(0^+_1)$, the delocalization of the $\alpha$ cluster is weak and 
gives a minor contribution to 
the total entanglement entropy.

The present result shows that the entanglement entropy is generated by the delocalization of clusters. 
The entanglement entropy is sensitive to the localization and delocalization of clusters and 
it can be a good measure to discuss the delocalization of clusters in 1D cluster states. 
In the present paper, I investigated the entanglement entropy only in 1D cluster wave functions, which are regarded as intrinsic states of 
3D nuclear systems, and discuss the delocalization in the 1D motion, i.e., the delocalization in
 radial motion of clusters. 
Zero entanglement entropy in 
the BB wave functions for localized cluster wave functions indicates that 
clusters are not delocalized but localized in the intrinsic systems. 
However, I should comment that, in realistic 3D wave functions, 
the 1D cluster wave functions should be projected on to the angular-momentum eigen states. 
The angular momentum projection to $0^+$ states from the 1D cluster wave functions usually 
causes the delocalization of clusters in angle motion. In that sense, the localized cluster wave functions 
may contain implicitly the delocalization in the angle motion even though 
the delocalization does not occur in the radial motion.
The localization of clusters in intrinsic states, for which 
the delocalization occurs in angle motion by the angular momentum projection,
is regarded as a kind of strong cluster correlations

In principle, the present method with entanglement entropy can be extended to 3D cluster systems, but 
it is practically not easy to calculate entanglement entropy in 3D because high dimensional single-particle 
bases are needed in diagolization of the one-body density matrix in 3D.
Moreover, as explained in Appendix \ref{app:3D}, 
the entanglement entropy  in 3D systems should have strong dependence on the single-particle angular momentum $j$ 
at the Fermi surface, because considerable entanglement entropy is generated 
in the angular momentum projection, 
which might wash out a pure contribution from the delocalization of clusters in the entanglement entropy.

\acknowledgments
The author would like to thank H.~Iida and T.~Suhara for helpful discussions.
The calculations of this work have been done using computers at Yukawa Institute for Theoretical Physics, Kyoto University.
This work was supported by 
JSPS KAKENHI Grant No. 26400270.

\appendix
\section{Density matrix}
I describe density matrices. For details of the one-body density matrix, the readers are referred to, for example, 
Ref.~\cite{ring-schuck}.   
An $A$-body density matrix for a wave function $\Psi(\bvec{r}\sigma,\bvec{r}_2\sigma_2,\ldots,\bvec{r}_A\sigma_A)$ for an $A$-nucleon system is
defined in the coordinate space as 
\begin{eqnarray}
&&\rho^{(A)}(\bvec{r}_1\sigma_1,\bvec{r}_2\sigma_2,\ldots,\bvec{r}_A\sigma_A; \bvec{r}'_1\sigma'_1,\bvec{r}'_2\sigma'_2,\ldots,\bvec{r}'_A\sigma'_A)\nonumber\\
&&=A!\Psi^*(\bvec{r}'_1\sigma'_1,\bvec{r}'_2\sigma'_2,\ldots,\bvec{r}'_A\sigma'_A) \Psi(\bvec{r}_1
\sigma_1,\bvec{r}_2\sigma_2,\ldots,\bvec{r}_A\sigma_A)\nonumber\\
&&= \langle\Psi| a^\dagger( \bvec{r}'_1\sigma'_1)   a^\dagger( \bvec{r}'_2\sigma'_2) \cdots   a^\dagger( \bvec{r}'_A\sigma'_A) 
 a^\dagger( \bvec{r}_A\sigma_A)\cdots
 a^\dagger( \bvec{r}_2\sigma_2)
 a^\dagger( \bvec{r}_1\sigma_1)  |\Psi\rangle,
\end{eqnarray}
where $a^\dagger(\bvec{r}\sigma)$ and $a(\bvec{r}\sigma)$ are 
creation and annihilation operators of a nucleon at the position $\bvec{r}$ 
with the spin-isospin $\sigma=p\uparrow,p\downarrow,n\uparrow,n\downarrow$.
$\Psi$ is normalized as $\langle \Psi|\Psi \rangle=1$.
$\rho^{(A)}$ is regarded as the matrix element of the $A$-body density operator 
$\hat\rho^{(A)}_{\Psi}=|\Psi\rangle\langle\Psi|$. 
The one-body density matrix is defined in the coordinate space as
\begin{equation}
\rho^{(1)}(\bvec{r}\sigma;\bvec{r}'\sigma')=\langle\Psi^{(A)}|a^\dagger(\bvec{r}'\sigma')
a(\bvec{r}\sigma) |\Psi^{(A)} \rangle.
\end{equation} 
$\rho^{(1)}$ is given by the trace of the  $A$-body density matrix 
\begin{eqnarray}
\rho^{(1)}(\bvec{r}\sigma;\bvec{r}'\sigma')&=&A\sum_{\sigma_2,\ldots,\sigma_A}\int d\bvec{r}_2 \ldots d\bvec{r}_A 
\Psi^{*}(\bvec{r}'\sigma',\bvec{r}_2\sigma_2,\ldots,\bvec{r}_A\sigma_A)\Psi(\bvec{r}\sigma,\bvec{r}_2\sigma_2,\ldots,\bvec{r}_A\sigma_A)\nonumber\\
&=&A {\rm Tr}_2 \cdots {\rm Tr}_A \rho^{(A)},
\end{eqnarray}
which is also called reduced density matrix.
The one-body density matrix is regarded as the matrix element of the one-body density 
operator $\hat \rho^{(1)}_{\Psi}$ for the wave function $\Psi^{(A)}$, 
\begin{equation}
\rho^{(1)}(\bvec{r}\sigma;\bvec{r}'\sigma')=\langle \bvec{r}\sigma|\hat \rho^{(1)}_{\Psi}|
\bvec{r}'\sigma'\rangle,
\end{equation} 
and I get 
\begin{eqnarray}
\rho^{(1)}(\bvec{r}\sigma;\bvec{r}'\sigma')&=&\sum_{pq} \varphi_p(\bvec{r}\sigma)
\rho^{(1)}_{pq}\varphi^*_q(\bvec{r}'\sigma')\nonumber\\
&=&\sum_{pq} \langle\bvec{r}\sigma|p\rangle
\rho^{(1)}_{pq}\langle q| \bvec{r}'\sigma' \rangle, 
\end{eqnarray} 
where 
\begin{equation}
\rho^{(1)}_{pq}=\langle\Psi^{(A)}|c^\dagger_qc_p|\Psi^{(A)}\rangle
\end{equation}
is the matrix element of the one-body density operator $\hat \rho^{(1)}_{\Psi}$ in arbitrary orthonormal bases.
$\hat \rho^{(1)}_{\Psi}$ is 
a Hermitian single-particle operator and it has the form
\begin{equation}
\hat \rho^{(1)}_{\Psi}=\sum_{pq}|p\rangle \rho^{(1)}_{pq} \langle q|. 
\end{equation}
The density matrix can be diagonalized by a unitary transformation of 
single-particle bases
\begin{eqnarray}
(D^\dagger \rho^{(1)} D)_{ll'}&=&\rho^{(1)}_l \delta_{ll'},\\
a^\dagger_l&=&\sum_{l'} D_{l'l} c^\dagger_{l'},
\end{eqnarray}
where
\begin{eqnarray}
\rho^{(1)}_l&=&\langle\Psi^{(A)}|a^\dagger_l a_l  |\Psi^{(A)}\rangle,\\
0&\leq& \rho^{(1)}_l \leq 1
\end{eqnarray}
is the eigen value of the density matrix and means the occupation number of 
the single-particle state $|l\rangle$ in the wave function $\Psi^{(A)}$.
In the coordinate space, the diagonal element 
$\rho^{(1)}(\bvec{r}\sigma)=\rho^{(1)}(\bvec{r}\sigma,\bvec{r}\sigma)$ of the density matrix is the one-body density 
of $\sigma=p\uparrow,p\downarrow,n\uparrow,n\downarrow$ nucleons at the position $\bvec{r}$,
and  it is expressed in the bases $|l\rangle$ as,
\begin{equation}
\rho^{(1)}(\bvec{r}\sigma)=\sum_l\rho^{(1)}_l\phi^*_l(\bvec{r}\sigma)\phi_l(\bvec{r}\sigma).
\end{equation}
The trace of the density matrix $\rho^{(1)}$ equals to the particle number:
\begin{equation}
A={\rm Tr} \rho^{(1)}=\sum_p\rho^{(1)}_{pp}=\sum_l\rho^{(1)}_l=\sum_\sigma \int\rho^{(1)}(\bvec{r}\sigma) d \bvec{r}.
\end{equation}

\section{Entanglement entropy} 

The von Neumann entropy of an $A$-particle system is defined 
by the $A$-body density matrix as
\begin{equation}
S^{(A)}=-{\rm Tr}\rho^{(A)} \log \rho^{(A)},
\end{equation}
and trivially $S^{(A)}=0$ for the pure state $|\Psi\rangle$ because 
$\left\{\rho^{(A)}\right\}^2=\rho^{(A)}$. 

The entanglement entropy is defined by the one-body density matrix as 
\begin{equation}
S^{(1)}=-{\rm Tr}\rho^{(1)}\log \rho^{(1)}= -\sum_l \rho^{(1)}_l \log \rho^{(1)}_l.
\end{equation}
Note that the entanglement entropy can be defined by the general 
$A'$-body density matrix for $A'\le A-1$ which is obtained by partial trace of 
the complete $A$-body density matrix for some degrees of freedom, but
in the present paper, I only consider the entanglement entropy for the one-body density
matrix.
The entanglement entropy is zero if and only if 
a wave function $|\Psi^{(A)}\rangle$ can be written by a Slater determinant because $\rho^{(1)}_l=1$ or 0
for a  Slater determinant as $\rho^{(1)}_l=1$ for occupied 
single-particle states and  $\rho^{(1)}_l=0$ for unoccupied states.
It is equivalent to the following theorem:
A wave function $|\Psi^{(A)}\rangle$ is a Slater determinant if
and only if the corresponding density operator $\hat \rho^{(1)}_{\Psi}$
satisfies $\{\hat \rho^{(1)}_{\Psi}\}^2=\hat \rho^{(1)}_{\Psi}$
 in the single-particle Hilbert space \cite{ring-schuck}.
It means that, for the ideal case of an uncorrelated Fermion system that 
the wave function is given by a Slater determinant, the system
has exactly zero entanglement entropy, $S^{(1)}=0$. $S^{(1)}$ is finite 
only if a system contains many-body correlations beyond the expression of 
a single Slater determinant. It  
indicates that the entanglement entropy can be an indicator for 
many-body correlations.

In the present paper, I define local entanglement entropy and 
analyze spatial distribution of the entanglement entropy. 
In analogy to the expression 
of the particle number by the $\sigma$ sum and 
$\bvec{r}$ integral of the density $\rho^{(1)}(\bvec{r}\sigma)$, 
I define the local entanglement entropy as follows, 
\begin{eqnarray}
S^{(1)}&=&  \sum_\sigma \int   s^{(1)}(\bvec{r}\sigma) d\bvec{r},\\
s^{(1)}(\bvec{r}\sigma)&=&\sum_l  \left[-\rho^{(1)}_l\log\rho^{(1)}_l\right]
 \phi^*_l(\bvec{r}\sigma)\phi_l(\bvec{r}\sigma).
\end{eqnarray}
$\phi^*_l(\bvec{r}\sigma)\phi_l(\bvec{r}\sigma)$ means the 
density distribution in the state $|l\rangle$ and normalized as $\sum_\sigma\int \phi^*_l(\bvec{r}\sigma)\phi_l(\bvec{r}\sigma)d\bvec{r}=1$, and
the factor $[-\rho^{(1)}_l\log\rho^{(1)}_l]$ is the contribution 
of the state $|l\rangle$ in the total entanglement entropy $S^{(1)}$.
Therefore, $s^{(1)}(\bvec{r}\sigma)$ reflects 
spatial distributions of the important states $|l\rangle$ that contribute to 
the total entanglement entropy. Note that $s(z)$ is not quantity 
determined only by local information at the position $z$.

\section{Entanglement entropy for correlated and uncorrelated systems in a toy model}
Let us consider correlated and uncorrelated wave functions 
in a simple toy model of an $A$ particle system.
For simplicity, $A$ particles are assumed to be 
distinguishable and stay on sites in a space. The number of available sites (single-particle states) is $N_s$
and I use the label $k_j$ for the $j$th single-particle states. Each particle can occupy 
one of  single-particle states, $k_j$ ($j=1,\ldots,N_s$).
A wave function $\Psi$ of an $A$-body state is expressed by a linear combination of 
direct products as 
\begin{equation}
\Psi(1,2,\ldots,A)=\sum_{\alpha_1}\sum_{\alpha_2}\cdots\sum_{\alpha_{A}}
C(\alpha_1,\alpha_2,\ldots,\alpha_A) \phi_{\alpha_1}(1) \phi_{\alpha_2}(2) \cdots  \phi_{\alpha_A}(A),
\end{equation}
where ${\alpha_i}$=$k_1,k_2,\ldots,k_{N_s}$, and $C(\alpha_1,\ldots,\alpha_A)$ 
is normalized as  $|\langle\Psi(1,2,\ldots,A)|\Psi(1,2,\ldots,A)\rangle|^2=1$.

Let us first consider uncorrelated systems.
If a state is an ideal state of independent particles, the wave function can be written by a simple product
of single-particle wave functions 
\begin{eqnarray}
\Psi(1,2,\ldots,A)=\psi_1(1)\psi_2(2)\cdots\psi_A(A),\\
\psi_i(i)=\sum_{\alpha=k_1,k_2,\ldots,k_{N_s}} c_i(\alpha)  \phi_{\alpha}(i)
\end{eqnarray}
and I get $S^{(1),i}=0$ 
because the one-body density operator for the $i$th particle is given as 
$\hat\rho^{(1),i}_{\Psi}=|\psi_i\rangle \langle \psi_i|$ and obviously satisfies 
$\{\hat\rho^{(1),i}_{\Psi}\}^2=\hat\rho^{(1),i}_{\Psi}$. 
(Here $\rho^{(1),i}$ is the reduced one-body density matrix defined
in subspace for the $i$th particle, and the entanglement entropy $S^{(1),i}$ is defined by $\rho^{(1),i}$.) 
Let us consider the state of 
free particles in zero momentum 
that all particles move freely in the 
whole system with an equal weight. The
wave function is given as 
\begin{equation}
\Psi(1,2,\ldots,A)=\frac{1}{N_s^{A/2}}\prod_{i=1}^{A} \left[\phi_{k_1} (i)+\phi_{k_2} (i)+\cdots+\phi_{k_{N_s}} (i) \right].
\end{equation}
The wave function has $S^{(1),i}=0$. Another example is  a ``localized cluster'' system of a cluster, where 
all particles are localized at one site $k_j$ to form a composite particle (a cluster) at 
 $k_j$. The wave function is given as
\begin{equation}
\Psi(1,2,\ldots,A)=\prod_{i=1}^{A} \phi_{k_j} (i).
\end{equation}
This localized cluster wave function also has zero entanglement entropy, 
$S^{(1),i}=0$.

Let us next consider the following example of a strong correlation limit, 
\begin{equation}
\Psi(1,2,\ldots,A)=\frac{1}{\sqrt{N_s}}\left\{
\prod_{i=1}^{A} \phi_{k_1} (i)+\prod_{i=1}^{A} \phi_{k_2} (i)+\cdots
\prod_{i=1}^{A} \phi_{k_{N_s}} (i)\right\},
\end{equation}
where $A$ particles form a composite particle, and the composite particle moves 
freely in the whole  space. This is a strongly correlated system, where, if a particle is observed 
at a certain site, all other particles are always observed at the same site. This is a strong coupling limit 
of the spatial correlation and corresponds to a delocalized cluster wave function of a cluster.
I can easily get the one-body density operator  
\begin{equation}
\hat \rho^{(1),i}= \sum_{j=1}^{N_s} \frac{1}{N_s}|k_j\rangle \langle k_j|,
\end{equation}
and the entanglement entropy
$S^{(1),i}=\log N_s$. 
 
Thus, the entanglement entropy indicates how a single particle is entangled with other particles.  
A localized composite particle
has $S^{(1),i}=0$ because there is no entanglement even though particles 
have some spatial correlation in a sense. 
In contrast,  if the delocalization of a composite particle occurs, $S^{(1),i}$ becomes
finite and  it is proportional to the logarithm of the number of 
sites (single-particle states) where the delocalization occurs. It indicates that the entanglement emerges
because of the delocalization of the composite parcle.

\section{Calculation of density matrix for linear $n\alpha$ states} 
In the present paper, I calculate matrix elements of the one-body density operator
for linear-chain $\alpha$-cluster states by the expansion of
localized Gaussian bases, 
\begin{equation}
\phi^{0s}_{R_k}=(\pi b^2)^{-3/4}\exp\left[ -\frac{1}{2b^2}(\bvec{r}-\bvec{R}_k)^2\right],
\end{equation}
with $\bvec{R}_k=(0,0,R_k)$. 
For simplicity, I choose a species of nucleons, for instance, $p\uparrow$ 
and describe only the spatial part of $\sigma=p\uparrow$  nucleons because
an $n\alpha$-cluster state is spin-isospin symmetric.
I take $R_k$ with 
0.75 fm intervals as $R_k=0.75 j$ fm ($j=0,\pm 1, \ldots, 15$), and prepare an orthonormal basis 
set $\{\phi_p(\bvec{r})\}$ from the non-orthonormal Gaussian bases 
$\{\phi^{0s}_{R_k}(\bvec{r})\}$ ($k=1,\ldots,35$). 
For the bases $|p\rangle$,
the one-body density matrix is written as 
\begin{eqnarray}
\rho_{pq}&=&\langle\Psi^{(A)}|c^\dagger_q c_p|\Psi^{(A)}\rangle,\nonumber\\
&=&\int d\bvec{r}d\bvec{r}' \phi^*_p(\bvec{r}) \rho^{(1)}(\bvec{r};\bvec{r}') \phi_q(\bvec{r}'),\\
\phi_p(\bvec{r})&=&\langle \bvec{r}|p\rangle.
\end{eqnarray}
By the diagonalization of $\rho_{pq}$, I can get the diagonalizing bases $|l\rangle$ by 
the unitary transformation of $|p\rangle$.

\section{3D-THSR wave functions of 2$\alpha$}\label{app:3D-THSR}
The deformed 3D-THSR wave function proposed by Funaki {\it et al.} \cite{Funaki:2002fn} is given as
\begin{eqnarray}\label{eq:3D-dTHSR}
\Phi^{2\alpha}_\textrm{3D-dTHSR}(\beta_\perp,\beta_z)&=&\int d\bvec{R}_1 d\bvec{R}_2\exp\left[
-\sum_{i=1,2}\left\{\frac{R^2_{ix}+R^2_{iy}}{\beta_\perp^2} -\frac{R^2_{1z}}{\beta_z^2}\right\}
\right]\Phi^{2\alpha}_{\rm BB}(\bvec{R}_1,\bvec{R}_2).
\end{eqnarray}
The spherical 3D-THSR wave function of the case $\beta_\perp=\beta_z=\beta$ 
is written as
\begin{eqnarray}\label{eq:3D-sTHSR}
\Phi^{2\alpha}_\textrm{3D-THSR}(\beta)&=&
\int d\bvec{R}_1 d\bvec{R}_2\exp\left[
-\sum_{i=1,2}\left\{\frac{\bvec{R}_i^2}{\beta^2}\right\}
\right]\Phi^{2\alpha}_{\rm BB}(\bvec{R}_1,\bvec{R}_2).\\
&\propto& {\cal A} \left[
\prod_{i=1,2}\exp\left\{-\frac{\bvec{X_i}^2}{\beta^2+b^2/2}
\right\}\phi(\alpha_i) \right],\nonumber\\
&\propto& {\cal A} \left[  \Phi_G(\beta)
\exp\left\{-\frac{\bvec{X}^2}{2\beta^2+b^2}\right\}
\phi(\alpha_1)\phi(\alpha_2)\right],\nonumber\\
\Phi_G(\beta)&\propto&\exp\left\{-\frac{2\bvec{X}_G^2}{\beta^2+b^2/2}\right\}
\end{eqnarray}
where $\bvec{X}_G=(\bvec{X}_1+\bvec{X}_2)/2$ and $\bvec{X}=\bvec{X}_1-\bvec{X}_2$,
and $\Phi_G(\beta)$ is the c. m. motion. The wave function can be rewritten as
\begin{eqnarray}
\Phi^{2\alpha}_\textrm{3D-THSR}(\beta)\propto\frac{\Phi_G(\beta)}{\Phi_G(0)}
\int d\bvec{R}\exp\left[-\frac{\bvec{R}^2}{2\beta^2}\right]
\Phi^{2\alpha}_{\rm BB}(\bvec{R}_1=+\frac{\bvec{R}}{2},\bvec{R}_2=-\frac{\bvec{R}}{2}).
\end{eqnarray}

\section{Extension to 3D system}\label{app:3D}
The analysis with the entanglement entropy is applied to 1D cluster systems in the 
present paper. In principle, it is able to calculate entanglement entropy  
also in 3D systems. However, to extract information of 
correlations from the entanglement entropy of 3D systems, 
one may encounter a problem from trivial 
correlation due to the total angular momentum projection.
Let us consider two particles (not identical to each other) 
in a spatial orbital with the angular momentum $l$.
The $L=0$ state after the total angular momentum projection is given 
as $\sum_{\mu}\frac{1}{2l+1}|l,\mu\rangle\otimes |l, -\mu\rangle$ and it 
has the finite entanglement entropy $S=\log(2l+1)$. 
In the case that a cluster develops spatially, 
nucleons in a cluster have strong spatial correlations, which are 
generally characterized by the mixing of high $l$ configurations,
and therefore, $S$ may reflect the many-body correlation 
in the cluster.
However, one should take care that, even for a $0^+$ state with a single $j^2$ 
configuration in the $jj$ coupling shell model the 
entanglement entropy is finite as $S=\log(2j+1)$
because of the trivial correlation by the total angular momentum projection.
It means that the entanglement entropy strongly depends on $j$ of the major shell
and such a large contribution from the angular momentum projection 
could make it difficult to
extract information of pure correlations beyond the $jj$ coupling 
configuration from  the entanglement entropy.

\end{document}